\documentclass[11pt]{article}
\usepackage[T1]{fontenc}
\usepackage[utf8]{inputenc}
\usepackage{microtype}
\usepackage[preprint]{acl}
\usepackage{times}
\usepackage{latexsym}
\usepackage{graphicx}
\usepackage{booktabs}
\usepackage{enumitem}
\usepackage{amsmath, amssymb}
\usepackage{tikz}
\usetikzlibrary{positioning,arrows.meta,shapes.geometric,fit,backgrounds,calc}
\usepackage{pgfplots}
\pgfplotsset{compat=1.18}
\setlist[itemize]{topsep=2pt, partopsep=0pt, itemsep=1pt, parsep=0pt}
\setlist[enumerate]{topsep=2pt, partopsep=0pt, itemsep=1pt, parsep=0pt}

\title{Ghost Tool Calls: Issue-Time Privacy for Speculative Agent Tools}

\author{Bardia Mohammadi$^{*}$ \quad Lars Klein$^{\dagger}$ \quad Akhil Arora$^{\ddagger}$ \quad Laurent Bindschaedler$^{*}$ \\[4pt]
  $^{*}$Max Planck Institute for Software Systems \quad $^{\dagger}$EPFL \quad $^{\ddagger}$Aarhus University \\[2pt]
  \texttt{bmohammadi@mpi-sws.org, lars.klein@epfl.ch}\\
  \texttt{akhil.arora@cs.au.dk, bindsch@mpi-sws.org}}

\date{}

\begin{document}
\maketitle

\begin{abstract}
Tool-augmented language agents speculatively issue likely future tool calls to hide latency, but those calls leak inferred user intent to external services before the agent commits to the branch. Every external observer that received the call retains the disclosure after the agent abandons the branch. Timing is the issue, not authorization: no commit-time cleanup, read-only restriction, or access-control allow-list unsends what an observer already holds. We call these invocations \emph{ghost tool calls} and propose Speculative Tool Privacy Contracts, a runtime abstraction that treats observation before commitment as a first-class effect, distinct from state mutation. We implement the contracts in a prototype runtime and evaluate twelve policies across three corpora. Speculative dispatch increases what an observer can infer about user intent; post-hoc filters, read-only restrictions, and access-control allow-lists leave that inference intact; only issue-time policies that change or suppress the speculative call's argument or destination projection before dispatch reduce it.
\end{abstract}

\section{Introduction}
\label{sec:intro}

Tool-augmented language agents act through external systems. They search the web, retrieve enterprise documents, inspect calendars, query CRMs, send messages, and call application APIs. Each call creates a privacy surface beyond the final answer: fact, timing, destination, and content of the call can reveal information about the user or organization.

\begin{figure}[t]
\centering
\includegraphics[width=\columnwidth]{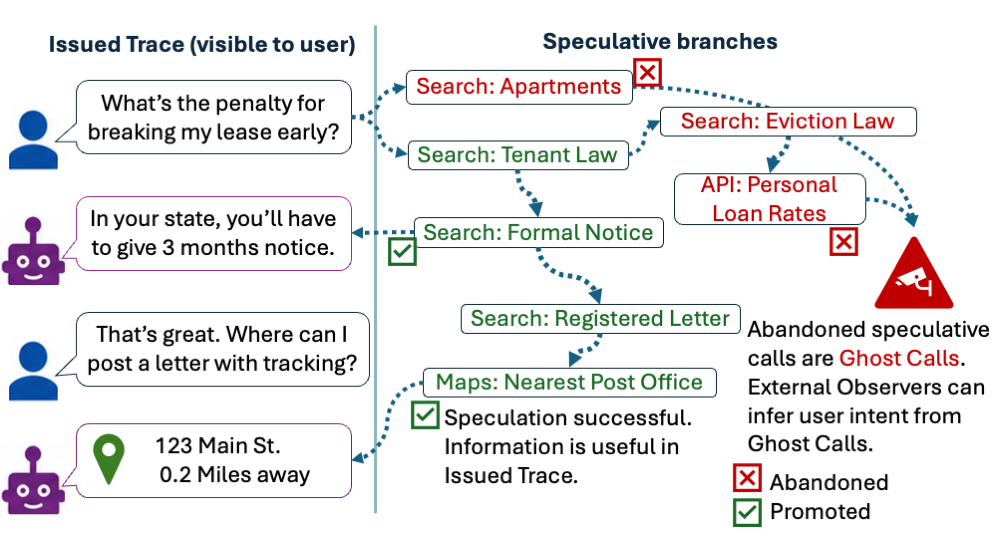}
\caption{Motivating trace: user-facing two-turn dialogue (left, what the user sees) and the runtime's speculative tree of dispatched tool calls (right). Green branches feed the answer; red branches are abandoned ghost calls.}
\label{fig:hero}
\end{figure}

Speculative tool execution widens that surface. A user might ask, ``What's the penalty for breaking my lease early?'' (Figure~\ref{fig:hero}). A careful agent answers the question. A latency-optimized agent fans out a tree of speculative queries: a search for tenant law, which the agent eventually uses; a sideways search for free apartments; and a deeper chain through eviction law into a personal-loan-rates API. The final reply uses the tenant-law search. The free-apartment and loan-rate branches do not, but the receiving providers have already logged them. The chain also encodes an inference the user never stated: the agent guessed the user might be looking to leave, might be facing eviction, and might need a loan to cover it. Speculation can persist across turns. A speculative pre-fetch about registered letters, issued while answering turn one, becomes the basis for the user's follow-up about mailing a tracked letter. The user's transcript looks innocuous; the provider's view contains a biography.

The mismatch lives in the trace. A speculative agent dispatches more tool calls than it ultimately uses, including calls along hypothetical branches the agent later abandons. We call those abandoned externally observable invocations \emph{ghost tool calls}: the runtime issued them, did not promote them to the committed plan, and cannot retract them. Each ghost call deposits \emph{observer-visible state} (provider logs, caches, audit sinks) from which an external observer can recover \emph{agent-inferred intent}: the topics, entities, destinations, and follow-up actions the runtime predicted the user might need.

Standard unsafe-tool framing misses this temporal mismatch. The same external search makes sense once the user has explicitly asked for legal resources; the runtime that ran it speculatively did so before knowing which workflow the agent would select. Access control answers whether the agent \emph{may} call a tool; speculative privacy asks whether the agent may reveal a tentative branch, destination, or argument before commitment. Speculative disclosure is distinct from parallel tool execution, where every call belongs to the selected plan.

Speculation is attractive because synchronous tool loops are slow, and recent systems expand speculation at the tool boundary~\cite{Sui2026ActWhileThinking,Ye2026SpeculativeActions,Guan2026DynamicSpeculativePlanning,Hooper2026SpeculativeInteractionAgents,Nichols2025SpeculativeToolCalls,Xia2026ToolSpec,Feng2026AsyncFC}. Existing safety abstractions reason about state mutation (read/write, reversible/irreversible) or about which committed calls a policy allows; none models the observer state created by issuing a call before the agent commits to needing it. The issue-time principle follows: for externally observable tools, the runtime must decide privacy policy before a speculative call leaves the trusted boundary, since commit-time cleanup has no privacy semantics once an observer has recorded the call.

We propose Speculative Tool Privacy Contracts to enforce this principle. A contract labels a proposed event by execution mode, observer, sensitivity, destination, confidence, and audit policy. Before external dispatch, the monitor decides whether the call issues unchanged, transforms, routes to a trusted local substitute, defers until branch promotion, or blocks. The design treats observation before commitment as an effect distinct from read/write state mutation.

We are not aware of prior work that formulates or evaluates privacy policy at the speculative-issue boundary. The closest existing baseline is Speculative Interaction Agents (SIA)~\cite{Hooper2026SpeculativeInteractionAgents}, which restricts speculation to read-only tools, but its restriction does not transform what reaches the observer and does not change the inference an adversary can draw from the trace.

We make three contributions:
\begin{itemize}[leftmargin=1.5em,itemsep=0em]
    \item The ghost-tool-call concept and the issue-time vs commit-time distinction, with a trace model that compares the issued trace against a committed-only counterfactual.
    \item Speculative Tool Privacy Contracts as an issue-time runtime abstraction with concrete labels, five actions, and a soundness invariant.
    \item An empirical comparison of 12 policies across three generated corpora plus a 66-task AgentDojo external-validity subset, with prototype runtime, frozen frontiers, and multiple observer adversaries to be released for verification.
\end{itemize}

\section{Background and Motivation}
\label{sec:background}

Tool use turns language agents into distributed systems. A single user request can pull in model services, data systems, user-context stores, and enterprise APIs. In the standard synchronous loop, the model proposes a tool call, the runtime executes it, and the model waits for the result before continuing. The agent walks a committed trace, one call at a time, and the loop is simple but slow when tools have network latency or a task needs many calls.

\paragraph{From slow loops to speculation.}
The standard escape from a slow sequential loop is lookahead. CPUs hide pipeline stalls with branch prediction; storage systems hide I/O latency with prefetching; agent runtimes do both at the tool boundary, predicting likely next calls and issuing them while the planner is still deciding which branch to take. Lookahead at the tool boundary is \emph{speculative tool execution}. The predicted invocations are \emph{speculative tool calls}: tool invocations issued before the agent has selected the branch that requires them. Concrete forms include (i) prefetching likely search or retrieval results, (ii) launching predicted API calls in parallel, (iii) resolving candidate entities before the final plan is known, and (iv) issuing duplicate or alternative calls before a previous call has failed. The defining property is abandonment risk: parallel calls all belong to a committed plan, whereas speculative calls may later be abandoned.

\paragraph{Adjacent optimizations.}
Recent tool-calling systems expand speculation in different places. PASTE, Speculative Actions, Dynamic and Interactive Speculative Planning, and Speculative Interaction Agents broaden the lookahead horizon at the tool boundary~\cite{Sui2026ActWhileThinking,Ye2026SpeculativeActions,Guan2026DynamicSpeculativePlanning,Hua2025InteractiveSpeculativePlanning,Hooper2026SpeculativeInteractionAgents}; in-engine token-level drafting of tool calls or schemas adds speculation inside the model-serving layer~\cite{Nichols2025SpeculativeToolCalls,Xia2026ToolSpec}; future-based asynchronous function calling overlaps decoding and execution whenever dependencies permit~\cite{Feng2026AsyncFC}; transactional systems handle consistency of state-changing speculative effects~\cite{Mohammadi2026Atomix}. Each reasons about correctness, idempotence, or state-update equivalence on the chosen plan, not about what each speculative call reveals to the observer that receives it. Our work adds a privacy axis to whatever speculative regime a runtime adopts. We do not argue against speculation; we argue that speculation needs an issue-time policy on what crosses the runtime boundary.

\paragraph{The missing effect.}
Existing abstractions distinguish safe from unsafe speculative calls by task-state mutation: when the losing branch leaves task state unchanged, the runtime discards it. Privacy needs a separate observation boundary. The observer-visible state from \S\ref{sec:intro} persists at each external observer regardless of whether the branch enters the committed plan.

\section{Ghost Calls and the Privacy Boundary}
\label{sec:model}

Let an agent runtime maintain candidate branches and label each tool event $e$ before dispatch. The \emph{issued trace} $S$ is the ordered log of every dispatched tool event; the \emph{committed trace} $C \subseteq S$ is the ordered subsequence whose results or side effects the runtime ultimately uses. A \emph{speculative frontier} is the set of candidate calls issued before the runtime has committed to the branch requiring them. A speculative call is \emph{promoted} if its result becomes part of $C$ and \emph{abandoned} otherwise. A \emph{ghost tool call} is an abandoned speculative event with a nonempty externally observable projection.

\paragraph{Disclosure channels.}
\label{sec:channels}
A tool event can disclose through three channels. \emph{Arguments}, the literal payload of the call, often carry the sensitive fact directly: a speculative search for ``tenant rights lease termination penalty'' reveals the user's frame to the provider whether or not the result is used. \emph{Destination}, the endpoint, host, or collection, can disclose topical intent independent of payload: a retrieval routed to an HR-leave, oncology, immigration, or security collection encodes the agent's interpretation of the task even with an empty query, so collection layout and access-control design become part of the privacy surface. \emph{Metadata}, including tool name, sizes, order, timing, retry counts, cache state, and branch traces, is deployment-sensitive but exploitable when destinations discriminate, caches are shared, or logs preserve branch status.

\paragraph{Observer projections.}
For observer class $i$, write $O_i(\cdot)$ for the projection of a trace into the channels observer $i$ can recover; different observers preserve different subsets (Table~\ref{tab:observers}). Then $O_i(S)$ is what $i$ actually records, and $O_i(C)$ is the \emph{committed-only counterfactual}: the projection $i$ would have seen had the runtime issued only $C$ from the start, with no pre-commit speculation. (This is not $S$ restricted to $C$: speculative events perturb provider-side timing, cache state, retry counters, and rate-limit ledgers.) The \emph{marginal exposure} is $\Delta_i = O_i(S) \setminus O_i(C)$: the ordered multiset of (arguments, destination, metadata) tuples and field deltas observer $i$ sees in $S$ but not in $C$. When the question is semantic rather than field-level, we measure \emph{inference advantage}: the increase in an adversary's probability of recovering a sensitive label given $O_i(S)$ versus $O_i(C)$.

\begin{table}[t]
\centering
\small
\begin{tabular}{@{}p{0.18\linewidth}p{0.72\linewidth}@{}}
\toprule
Observer & Typical visibility \\
\midrule
Provider & Tool, endpoint, tenant, arguments, timestamps, sizes, provider logs. \\
Network & Destination, timing, sizes, retries, traffic shape (TLS hides plaintext args). \\
Runtime log & Full trace including branch labels, transformed arguments, cache keys. \\
Co-tenant & Cache occupancy, lock contention, queueing delay, rate-limit residue. \\
Auditor & Whatever the runtime records. \\
\bottomrule
\end{tabular}
\caption{Observer projections used in the trace model. Each row lists fields that observer class can recover.}
\label{tab:observers}
\end{table}

\paragraph{Promoted versus abandoned.}
\emph{Ghost calls} (abandoned speculative events) lie in $S \setminus C$ by construction, so their projections contribute to $\Delta_i$ whenever observer $i$ records anything about them. Promoted speculative events appear once on the wire whether the agent speculated or not, so they appear in $\Delta_i$ only against observers that preserve ordering or timing: a provider logging only canonical arguments and destination receives the same record as in the no-speculation counterfactual, while a provider that timestamps requests records the speculative-phase timestamp instead of the would-be commit-phase one. Throughout, \emph{ghost tool call} names the strict abandoned-event case; \emph{pre-commit disclosure} names the broader category that also includes promoted speculative events under timing- or ordering-sensitive observers. The monitor in \S\ref{sec:contracts} evaluates every speculative event at issue time regardless of later promotion, because disclosure happens at issue time and the runtime cannot retract it. A worked trace with $S$, $C$, $O_i(S)$, $O_i(C)$, and $\Delta_i$ is in Appendix~\ref{app:worked-trace}.

\paragraph{Proposition 1 (No commit-time erasure).}
\emph{Assume observer $i$ persists observations outside the runtime's commit boundary. If a ghost call $g$ emits a nonempty projection $O_i(g)$ at issue time, no policy that executes only after branch commitment can remove that projection from observer $i$'s prior view.} Once a provider, network observer, shared cache, or audit log has received the external trace of $g$, the runtime cannot retroactively unsend it. Rollback works for state effects because a runtime can delay or compensate writes~\cite{Mohammadi2026Atomix}; observation effects sit outside that boundary. Issue-time is therefore the only enforcement point for externally observable speculative dispatch.

\paragraph{Proposition 2 (Read-only is not observation-free).}
\emph{Read-only authorization does not imply privacy preservation under speculation: read-only ghost calls can emit nonempty observer projections through any of the three channels above.} Designs that permit speculation only for read-only tools~\cite{Hooper2026SpeculativeInteractionAgents} protect against state mutation, not observation.

\section{Speculative Tool Privacy Contracts}
\label{sec:contracts}

Speculative Tool Privacy Contracts make the \emph{observer-visible-before-commit} effect explicit and attach policy decisions to it.

\begin{table}[t]
\centering
\small
\begin{tabular}{@{}p{0.27\linewidth}p{0.63\linewidth}@{}}
\toprule
Field & Meaning \\
\midrule
Execution mode & committed, speculative, prefetch, retry, validation, shadow. \\
Effect class & pure-local, local-read, external-read, external-write, semi-stateful, irreversible. \\
Observer class & provider, network, runtime log, co-tenant, auditor. \\
Argument labels & public, personal, sensitive, intent-revealing. \\
Destination label & public, tenant-internal, sensitive, regulated, forbidden. \\
Branch confidence / budget & calibrated commit estimate; cumulative per-task disclosure budget. \\
Audit policy & which observer (if any) receives the audit record of this event. \\
\bottomrule
\end{tabular}
\caption{Fields the monitor binds to each speculative event before dispatch.}
\label{tab:contract-fields}
\end{table}

\paragraph{Why a multi-field vocabulary and not a scalar risk level?}
Collapsing the vocabulary to a single (mode, risk-level) pair would force every policy to encode its decision inside a policy-specific risk function. That relocates the complexity. Different policies read different fields: Rewrite reads argument sensitivity, Shadow reads destination, Gate reads branch confidence, Drop reads sensitivity anywhere. Sensitivity and confidence are also orthogonal in our measurements (the Gate lever in Table~\ref{tab:gate-sweep} lives in the high-sensitivity, low-confidence quadrant a scalar would flatten), and one destination yields multiple observer projections, so policies must track $\Delta_i$ per observer.

\paragraph{Trusted boundary.}
The monitor sits between the planner and \emph{every} externally observable IO path. Tool providers, network infrastructure, shared caches, audit sinks, retry queues, telemetry exporters, tracing spans, rate-limit ledgers, and auth-token minting all sit \emph{behind} the monitor: raw speculative arguments must not enter logs, queues, metrics, traces, or caches before a policy decision. The Trusted Computing Base (TCB) contains the planner, the speculation controller, the contract monitor, and the dispatcher up to external dispatch. Our threat model trusts the planner and leaves model-provider mediation to future work (Limitations).

\paragraph{Decision function.}
For a proposed event $e$ under contract $\kappa$, the monitor computes $D(e,\kappa) \to \{\textsc{allow}, \textsc{rewrite}, \textsc{shadow}, \textsc{defer}, \textsc{block}\}$ as a fail-closed pipeline (any missing label or uncovered case maps to \textsc{Defer} or \textsc{Block}, never silently to \textsc{Allow}): \textsc{Block} if the tool or destination is outside $\kappa$'s allow-list; \textsc{Defer} if the event is speculative and its branch confidence falls below $\kappa$'s threshold; \textsc{Rewrite} if any argument is labeled HIGH-sensitivity; \textsc{Shadow} if a non-provider-visible substitute is required and the destination is external; \textsc{Defer} if the per-task disclosure budget is exceeded; otherwise audit and \textsc{Allow}. The audit hook records the decision and transformed fields; raw speculative arguments never persist unless $\kappa$ explicitly grants that observer. Pseudocode in Appendix~\ref{app:mechanism}.

\paragraph{Actions.}
\textsc{Allow} dispatches the event unchanged.
\textsc{Rewrite} applies a declared transformation (entity redaction, time-range coarsening, intent-phrase replacement) before dispatch.
\textsc{Shadow} executes a trusted-local substitute and marks the result non-provider-visible. Our prototype reuses the substituted result if the runtime later promotes the speculative call to the committed plan, so substitute quality conditions downstream task success; a cleaner refinement would re-issue the call against the real provider on promotion, when the committed plan justifies the disclosure.
\textsc{Defer} queues the event for re-evaluation on branch promotion; abandoned branches terminate without external dispatch.
\textsc{Block} drops the event with a policy denial. Every event traverses four states with explicit guards (Issued $\to$ Decided $\to$ Dispatched $\to$ Completed); promotion of a previously \textsc{Deferred} event re-runs $D(e,\kappa)$ as a committed event before external dispatch.

\paragraph{Theorem 1 (Soundness under complete mediation).}
Assume (A1) complete mediation: every speculative or committed event reaches the dispatcher only through the monitor, and no telemetry, retry queue, cache, or audit sink receives raw event fields before the monitor decides; (A2) label correctness: labels correctly classify each event's destination, arguments, and effect class; (A3) action preservation: rewrites and shadow substitutes preserve their declared field-label restrictions (a Rewrite transforming an intent-revealing argument yields a generic argument that no longer carries the forbidden label; a Shadow substitute does not externally emit the forbidden field). \emph{Under} A1--A3, the monitor's evaluation of $D(e,\kappa)$ at issue time prevents any field of any label $\ell$ that $\kappa$ forbids at observer $i$ from entering $\Delta_i$, regardless of whether the event is later committed, abandoned, deferred, or promoted. Each of A1--A3 is necessary. Appendix~\ref{app:mechanism} formalizes the event schema, label lattice, and action semantics.

\section{Evaluation}
\label{sec:evaluation}

We ask whether speculative provider-visible traces support an issue-time privacy claim and which controls shrink the resulting marginal exposure $\Delta_i$. We measure adversarial recovery of a hidden sensitive label from observer projections, not final-answer correctness. We use three corpora plus an AgentDojo external-validity subset, with sizes stated per RQ; Claude Opus 4.7 is the primary adversary. Appendix~\ref{app:tables} gives the full tables.

\paragraph{Setup.}
The prototype is a Python contract dispatcher over five tool adapters (web search, document retrieval, calendar lookup, CRM lookup, email search), each carrying effect, argument-privacy, and destination-privacy labels. We swap in a Brave Search adapter for the real-provider study. Each closed-set adversarial inference picks one of the per-corpus labels from the projection it sees; we hold the speculative frontier byte-identical across paired replays and vary the dispatcher policy, so sign tests use the matched (task, seed) structure. The planner emits a median of 8 speculative candidates per task with mean ghost-call rate 82--89\% across corpora (Appendix~\ref{app:frontier-stats}). We organize the evidence around five research questions.

\paragraph{Policies and baselines.}
We compare twelve policies that share the same speculative frontier and differ only in dispatcher behavior. The \emph{endpoints} are \textbf{Naive} (no policy, every candidate dispatches) and \textbf{No-Spec} (committed plan only, no speculation). \emph{Post-hoc filters} act after a call has already issued: \textbf{SIA} (Speculative Interaction Agents) admits only read-only tools~\cite{Hooper2026SpeculativeInteractionAgents}; \textbf{ACL} gates dispatch by a per-tool allow-list; \textbf{Late-Scrub} scrubs the issued trace after branch commitment. \emph{Issue-time transformations} change what reaches the observer before dispatch. \textbf{Pre-Scrub} is an issue-time DLP-style weak baseline: it runs a free-text PII regex before issue, which leaves structured argument and destination fields untouched. \textbf{Rewrite} paraphrases sensitive arguments into generic equivalents; \textbf{Shadow} routes the call to an internal substitute whose result is non-provider-visible; \textbf{Gate} defers low-confidence speculative calls until branch selection; \textbf{Taint} blocks calls with substring matches to hand-curated taint sources; and \textbf{Taint-F} tracks taint at field granularity. The refusal baseline, \textbf{Drop}, rejects any call tagged sensitive. SIA is the only directly attributable prior speculative-agent baseline. ACL, Pre-Scrub, and Late-Scrub instantiate standard allow-listing, free-text redaction, and post-hoc transcript scrubbing; Taint and Taint-F adapt classical IFC ideas~\cite{Denning1976Lattice,Myers1999JFlow}; Naive and No-Spec are the speculative and non-speculative endpoints; Rewrite, Shadow, and Gate are contract actions.

\paragraph{RQ1: Do speculative traces leak inferred intent?}
Yes. The gap is large, and it survives moving from a static plan to a real LLM planner.

\emph{Mechanism.} 30 hand-crafted tasks with fixed committed plans, three seeds, Opus 4.7 adversary (Table~\ref{tab:headline}). Holding the committed plan constant isolates the ghost-call contribution: recovery climbs from 0.44 under No-Spec to 0.97 under Naive; Late-Scrub, SIA, and ACL track Naive at 0.97. Only Rewrite (0.53) and Shadow (0.42) move the metric.

\emph{Realism.} The same 30 tasks under a SpeculativeAgent LLM planner (three seeds; Table~\ref{tab:expanded-11}): Naive recovers 0.49, Shadow 0.00, Rewrite 0.04. The Naive-equivalent cluster persists; Rewrite and Shadow remain near the floor without a held-constant plan.

Speculative dispatch produces a large, mechanistically attributable leak; post-issue scrubs cannot retract it. Appendix~\ref{app:tables} reports per-policy CIs and call counts.

\paragraph{RQ2: Which controls reduce leakage?}
Only issue-time transforms. We rule out a planner-frontier confound by replaying one Naive-sampled planner trace per (task, seed) through every policy.

\begin{figure*}[t]
\centering
\begin{tikzpicture}
\begin{axis}[
  width=0.88\textwidth,
  height=6.7cm,
  xlabel={Paired $\Delta$ vs.\ Naive},
  ytick={1,2,3,4,5,6,7,8,9,10,11,12},
  yticklabels={Shadow, Rewrite, Taint, Taint-F, Drop, Gate, No-Spec, Late-Scrub, ACL, Naive (ref), Pre-Scrub, SIA},
  y dir=reverse,
  ymin=0.3, ymax=12.7,
  xmin=-0.16, xmax=0.07,
  xmajorgrids=true,
  ymajorgrids=false,
  major grid style={gray!30, dashed},
  tick label style={font=\footnotesize},
  label style={font=\footnotesize},
  legend style={font=\footnotesize, at={(0.02,0.02)}, anchor=south west},
]
\addplot[mark=none, black, thick, dashed] coordinates {(0,0.3) (0,12.7)};
\addplot+[
  only marks, mark=*, mark size=2pt, color=black,
  error bars/.cd, x dir=both, x explicit,
] coordinates {
  (-0.104, 1) +- (0.028, 0.029)
  (-0.076, 2) +- (0.028, 0.029)
  (-0.033, 4) +- (0.027, 0.024)
  (-0.027, 5) +- (0.023, 0.024)
  (+0.020, 12) +- (0.015, 0.018)
};
\addlegendentry{sign-test $p<0.05$}
\addplot+[
  only marks, mark=o, mark size=2.2pt, color=gray!70!black,
  error bars/.cd, x dir=both, x explicit,
] coordinates {
  (-0.027, 3) +- (0.026, 0.027)
  (-0.024, 6) +- (0.024, 0.024)
  (-0.024, 7) +- (0.023, 0.022)
  (+0.000, 8) +- (0.013, 0.011)
  (+0.000, 9) +- (0.013, 0.013)
  (+0.013, 11) +- (0.022, 0.022)
};
\addlegendentry{not significant}
\addplot+[only marks, mark=diamond*, mark size=3pt, color=black] coordinates {(0, 10)};
\end{axis}
\end{tikzpicture}
\caption{Paired-replay leakage by policy on 150 tasks and three seeds ($n{=}450$ per policy). Points show $\Delta$ in adversary recovery relative to Naive; negative values mean less leakage. Filled markers have sign-test $p{<}0.05$; the diamond marks Naive.}
\label{fig:forest}
\end{figure*}
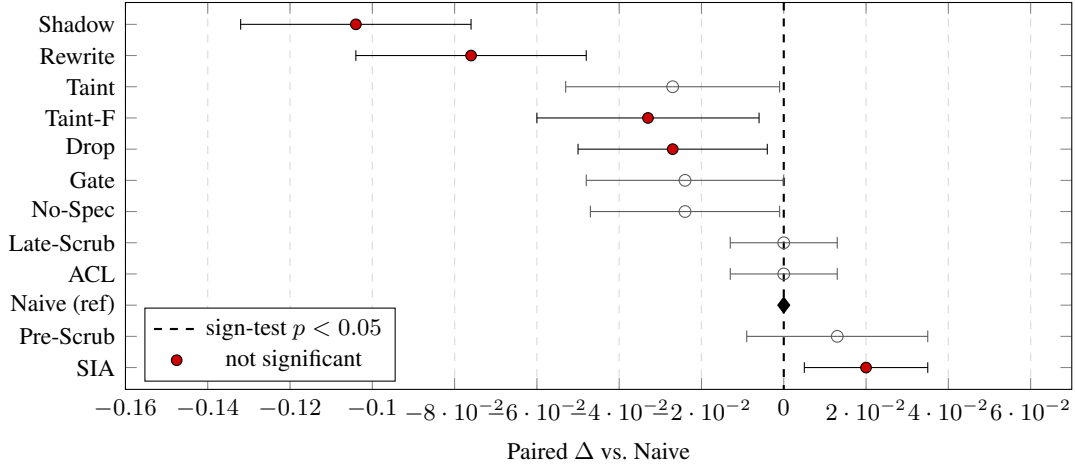

\emph{Paired-frontier replay.} Figure~\ref{fig:forest} uses 150 tasks $\times$ 3 seeds, or 450 paired trials per policy; Table~\ref{tab:paired-replay-450} gives the full numbers. The speculative frontier is identical across policies, and only the dispatcher decision changes. Four policies that do not change the structured projection (Pre-Scrub, Late-Scrub, ACL, SIA) cluster with Naive: Late-Scrub and ACL have zero delta; Pre-Scrub is statistically indistinguishable ($p{=}0.31$); SIA is the only policy with a statistically significant positive delta ($+0.02$, $p{=}0.04$), worse than Naive on this corpus because its read-only allow-list passes exactly the search/retrieve/lookup calls that carry the most intent signal. This is the leak: utility is preserved because the provider projection is preserved. Rewrite and Shadow reduce recovery at $p{<}0.0001$; a field-sensitive provenance variant (Taint-F) cuts the residual signal by $\approx 32\%$ (Naive 0.104 $\to$ Taint-F 0.071). We also implement a stronger IFC baseline modeled after FIDES~\cite{Costa2025Fides} with typed source declarations, a typed sink table, and field-level taint propagation: it reaches 0.131 recovery on the same corpus, worse than Rewrite (0.029) and Shadow (0.000), and within CI of Naive (0.104). The contract abstraction's transformation actions are not subsumed by careful IFC labeling.

\emph{Scale.} The pattern reproduces on the 150-task unpaired pooled corpus (Table~\ref{tab:pooled-150}). Absolute recovery drops as the candidate-label space grows from 30 to 150 (chance falls from $\approx$0.033 to $\approx$0.0067), but the four Naive-equivalent baselines stay within 0.02 of Naive, Rewrite at 0.02, Shadow at 0.00. The 150-task paired-replay Naive--No-Spec gap is smaller than the 30-task static gap because paired replay does not hold the committed plan constant; the static-plan benchmark, which does, attributes the gap mechanistically to ghost calls.

\emph{Label-noise robustness.} The soundness invariant rests on correct labels. False-negative noise that downgrades SENSITIVE/INTENT-REVEALING arguments and SENSITIVE destinations to PUBLIC at 5/10/20\% on the 150-task paired replay (Haiku adversary): at 20\%, Shadow at 0.08 and Rewrite at 0.06 remain significantly below noisy Naive, and the Naive-equivalent cluster stays within 0.01 spread. The ranking survives plausible classifier error; full sweep in Table~\ref{tab:label-noise-fn}.

Under byte-identical speculative input, post-issue scrubs do not move the metric; SIA preserves the same provider signal as Naive. The result holds at 5$\times$ scale and under 20\% label noise.

\paragraph{RQ3: What carries the residual signal?}
Destination URLs and structured argument fields, not tool names.

\emph{Tool-name masking.} Replacing every projection's tool name with a generic placeholder on the 30-task real-planner corpus produces zero delta for both Naive (0.49 with and without names) and Pre-Scrub (0.47 with and without); 25 of 30 per-task predictions match across conditions per seed. The residual signal lives in destination URL paths that literally name the topic and in structured argument fields whose values a free-text personally-identifiable-information (PII) regex does not cover.

\emph{Channel isolation.} A 9-task seed ablation drops one channel at a time under Naive (Table~\ref{tab:channels}): full projection 9/9 $\to$ args+names only 9/9 (argument dominates) $\to$ dest+names only 5/9 (destination contributes) $\to$ names only 3/9 (the floor).

\emph{Anonymized destinations.} On a seed-1 anonymization rerun of the 150-task paired-replay, neutralizing every destination URL shifts Naive recovery from 0.113 to 0.127 (paired CI overlaps zero) and leaves Rewrite/Shadow unchanged. The argument channel carries the leak on its own; topic-bearing destinations amplify rather than originate it. A 30-task enterprise-style corpus with anonymous endpoints by construction reproduces the ordering: Naive 1.00, Shadow 0.00, Rewrite 0.09 (Appendix~\ref{app:enterprise-corpus}).

A PII redactor that operates on free text but leaves structured destination and argument schemas untouched is, in our measurements, indistinguishable from no redaction. Effective controls must transform the structured projection (Rewrite) or replace the destination (Shadow).

\paragraph{RQ4: Do the findings hold under different adversaries and a real provider?}
Yes, on every axis we tested.

\emph{Cross-adversary.} Rescoring the same projections under three model families (Opus 4.7, Haiku-4.5, GPT-4o-mini; Table~\ref{tab:cross-adversary}) preserves the policy ranking within each model: Naive, Pre-Scrub, Late-Scrub, and ACL tie at the top; Shadow at 0.00 and Rewrite at 0.03 hold the floor. Pooled closed-set recovery is 0.33 (Opus), 0.25 (Haiku), 0.22 (GPT-4o-mini). Semantic open-set matching (sentence-embedding cosine) reaches 0.13 on Opus, roughly $5\times$ the strict-lexical floor. Non-LLM baselines (keyword 0.06, TF-IDF 0.08; Table~\ref{tab:non-llm-adversary}) trail LLM adversaries 4--5$\times$; a corpus leave-one-out Naive Bayes adversary trained on per-task labels reaches 0.65, an upper bound the LLM does not approach.

\emph{Metric robustness.} The policy ranking holds under top-$k$ accuracy ($k{=}1,3,5$; Spearman 0.83 against top-1), normalized mutual information, and an open-set semantic adversary on the 150-task corpus. Critically, a decoy-label adversary that sees the true label mixed with 10 unrelated decoys (chance $\approx 0.091$) recovers the true label at 0.90--0.93 on the Naive cluster — the adversary discriminates via projection content, not via candidate-list recognition. Shadow holds the floor under every alternative metric (Appendix~\ref{app:metric-expansion}).

\emph{Real provider.} A Brave Search sanity check (9 tasks, 1 seed) gives Naive 0.67 and Rewrite 0.44; a 30-task $\times$ 3-seed DuckDuckGo run reproduces the gap with non-overlapping 95\% CIs (Naive 0.49, Rewrite 0.04; Table~\ref{tab:ddg-inference}). The speculative search still happens, but the literal string reaching the provider under Rewrite is generic.

\emph{Async wall-clock.} Single-tool DuckDuckGo overlap saves 5.4\% at p50 (Table~\ref{tab:overlap-timing}); a multi-tool workload with $\ge 3$ parallel speculative calls saves 12.3\% at p50, with Contract-async (Rewrite, Gate, Taint-F) within 1.5~pp of Naive-async; Shadow is 2.7~pp slower because local substitutes shift the overlap window (Appendix~\ref{app:multi-tool-async}). Contract enforcement is not the latency bottleneck.

\emph{External validity.} 66 AgentDojo tasks (Table~\ref{tab:agentdojo-external}) reproduce the policy ordering qualitatively at one seed: the let-through baselines (Naive and SIA) tie at 0.75 and Pre-Scrub at 0.85 on privacy-bearing prompts, Taint at 0.13, and Rewrite and Shadow at 0.00. AgentDojo prompts name sensitive topics directly, so absolute numbers run high. The closed-set adversary in this table admits a \texttt{none} option, and under that menu Rewrite's neutral false-positive rate drops to 0.06 from a 1.000 closed-set-menu artifact reported in earlier runs (Appendix~\ref{app:agentdojo-metric-fix}).

The policy ordering holds across the adversaries we tested, reproduces at the HTTP boundary, and carries to a benchmark with a different threat profile; the contract is not the latency bottleneck.

\begin{table}[t]
\centering
\small
\begin{tabular}{@{}lccc@{}}
\toprule
Policy & Adv.\ recovery & FP rate & Calls \\
\midrule
Naive       & 0.750 & 0.327 & 5.02 \\
SIA\textsuperscript{$\dagger$} & 0.750 & 0.327 & 5.02 \\
Pre-Scrub   & 0.846 & 0.308 & 4.92 \\
Taint       & 0.130 & 0.019 & 0.32 \\
Rewrite     & 0.000 & 0.058 & 5.12 \\
Shadow      & 0.000 & 0.000 & 0.03 \\
\bottomrule
\end{tabular}
\caption{AgentDojo external-validity subset: 66 read-only tasks per policy (20 privacy-bearing + 46 neutral). Privacy-bearing inference rate and neutral-prompt false-positive rate under a Haiku-4.5 planner/adversary, with \texttt{none} admissible in the closed-set candidate menu. $\dagger$ SIA is read-only by construction on AgentDojo.}
\label{tab:agentdojo-external}
\end{table}

\paragraph{RQ5: What do non-provider observers reveal?}
Runtime logs reveal nearly as much as providers; network and shared-cache channels do not in this corpus. We measure three additional observer classes from Table~\ref{tab:observers} on the same 30-task corpus and 12 policies.

\emph{Runtime log.} An adversary that sees tool name, a 20-character per-value argument summary, the dispatcher decision tag, and a 100~ms duration bucket recovers 0.24 pooled, only 0.02 below the same model's full-projection provider adversary (0.26) on the same traces. Argument truncation is insufficient when summaries and decision tags survive.

\emph{Network only.} An adversary seeing host, port, payload size, and timing recovers 0.04, within sampling noise of the 0.033 random baseline. The traffic-shape claim in \S\ref{sec:channels} needs per-tenant or topic-bearing hostnames our corpus does not exercise.

\emph{Co-tenant cache.} A hand-built probe set with oracle access to candidate labels recovers 9/9 tasks; the same protocol learned by cross-validation on a one-task-per-label corpus recovers 0/9. The channel carries the sensitive label; exploitation needs a labeled prior overlapping the victim. Audit logs fold into runtime-log; network/shared-cache channels are deployment-sensitive.

\paragraph{Cost and Gate calibration.}
Naive uses 4.0 provider calls/task vs Shadow/No-Spec at 1.0 (static plans); the planner LLM, not the monitor, dominates wall-clock (Table~\ref{tab:runtime-latency}). Gate exposes a tunable threshold $\tau$ that walks a privacy/utility lever from 0.12 leakage at 5.4 calls/task (permissive $\tau{=}0.3$) to 0.06 at 0.96 calls/task (strict $\tau{=}0.9$) on the 150-task paired-replay (Table~\ref{tab:gate-sweep}).

\paragraph{Privacy/utility trade-off.}
We also measure task success with a Haiku-4.5 LLM-as-judge rubric on a 30-task stratified subset (10 health, 10 employment, 10 finance), all twelve policies, and three seeds (Table~\ref{tab:task-success-12}). Task-success clusters tightly in $[0.77, 0.83]$ with every CI overlapping every other, so the utility axis does not statistically separate the twelve policies; the privacy axis separates them from 0.00 (Shadow) to 0.26 (Naive, ACL). The Pareto frontier has two corners: Shadow at zero leakage, No-Spec at the utility ceiling but no speculative latency advantage (Table~\ref{tab:overlap-timing}). Gate at $\tau{=}0.7$ is a useful middle point (leakage 0.04, task-success 0.79). The post-hoc cluster (Late-Scrub, SIA, ACL) sits at or above Naive on leakage with no utility gain; speculative policies pay at most 0.04 task-success to bring leakage below 0.05.

\begin{table}[t]
\centering
\small
\setlength{\tabcolsep}{3pt}
\begin{tabular}{@{}lcc@{}}
\toprule
Policy & Adv.\ recovery & Task success \\
\midrule
\textbf{Shadow}          & \textbf{0.000 [0.000, 0.000]} & 0.794 [0.739, 0.850] \\
Rewrite                  & 0.033 [0.000, 0.078]          & 0.794 [0.744, 0.844] \\
Gate ($\tau{=}0.7$)      & 0.044 [0.011, 0.089]          & 0.794 [0.739, 0.850] \\
Drop                     & 0.067 [0.022, 0.122]          & 0.794 [0.744, 0.844] \\
\textbf{No-Spec}         & 0.067 [0.022, 0.122]          & \textbf{0.833 [0.783, 0.878]} \\
Taint                    & 0.100 [0.044, 0.167]          & 0.828 [0.778, 0.878] \\
Taint-F                  & 0.122 [0.067, 0.189]          & 0.800 [0.750, 0.850] \\
Pre-Scrub                & 0.211 [0.133, 0.311]          & 0.800 [0.750, 0.850] \\
Late-Scrub               & 0.233 [0.144, 0.322]          & 0.783 [0.733, 0.833] \\
SIA                      & 0.233 [0.144, 0.322]          & 0.778 [0.722, 0.828] \\
Naive                    & 0.256 [0.178, 0.344]          & 0.783 [0.733, 0.833] \\
ACL                      & 0.256 [0.167, 0.344]          & 0.772 [0.722, 0.822] \\
\bottomrule
\end{tabular}
\caption{Privacy/utility trade-off on a 30-task stratified subset, three seeds ($n{=}90$ per policy). Recovery is closed-set leakage; task success is a Haiku-4.5 judge score on final answers.}
\label{tab:task-success-12}
\end{table}

\section{Related Work}
\label{sec:related}

\paragraph{Speculative tool execution.}
Speculation at the tool boundary is established practice~\cite{Sui2026ActWhileThinking,Ye2026SpeculativeActions,Guan2026DynamicSpeculativePlanning,Hua2025InteractiveSpeculativePlanning,Hooper2026SpeculativeInteractionAgents,Nichols2025SpeculativeToolCalls,Xia2026ToolSpec,Feng2026AsyncFC,Mohammadi2026Atomix}; we add an issue-time privacy abstraction on top, not speculation itself. SIA is the closest issue-time baseline; Proposition~2 rules out its read-only restriction. These runtimes already balance token-and-API cost against latency; encoding privacy as cost throttles sensitive-call volume but does not change the projection of issued calls. \textsc{Rewrite} and \textsc{Shadow} are orthogonal: they transform per-call projection.

\paragraph{Agent IFC, guardrails, and side channels.}
FIDES~\cite{Costa2025Fides}, NeuroTaint~\cite{Cai2026NeuroTaint}, RTBAS~\cite{Zhong2025RTBAS}, and verifiably safe tool-use~\cite{Mou2026ToolSafe,Doshi2026SafeToolUse} gate selected committed tool calls; classical IFC and effect systems~\cite{Denning1976Lattice,Myers1999JFlow} label data and operations. Contracts add observation-before-commitment as a first-class effect over the speculative frontier. Adjacent side-channel work~\cite{Zhang2025AgentTrafficFingerprint,Wei2025SpeculationSpills,Kocher2019Spectre} studies different objects but shares the Spectre lesson: observations on uncommitted paths escape the intended execution model. Committed-leakage benchmarks~\cite{Huang2026CIPL,ElYagoubi2026AgentLeak,Debenedetti2024AgentDojo,Ruan2024ToolEmu,Zhan2024InjecAgent,Yao2025TauBench} measure attacks on the selected plan, not marginal speculative exposure.

\paragraph{Query privacy and contextual integrity.}
PIR~\cite{Chor1998PIR,Howe2008Trackmenot} and RAG privacy~\cite{Zeng2024RAGPrivacy} reduce what a service learns from a query; we ask whether to issue it. Contextual integrity~\cite{Nissenbaum2004ContextualIntegrity} captures the timing-and-purpose distinction.

\section{Conclusion}
\label{sec:conclusion}

Speculative tool execution creates an issue-time privacy failure: a not-yet-needed call discloses intent before commitment. Speculative Tool Privacy Contracts treat observation-before-commit as a first-class effect; only issue-time controls suppress it.

\section*{Limitations}
\label{sec:limits}

\paragraph{Complete mediation.}
Speculative Tool Privacy Contracts rely on complete mediation: any client library, cache, retry path, metric, trace, audit sink, or credential path that receives raw speculative fields before the monitor runs violates the model and the soundness invariant no longer applies.

\paragraph{Label correctness.}
The policy depends on labels and confidence estimates; missing annotations and classifier failures fail closed, trading utility for privacy. Mislabeling an intent-revealing argument as benign, or an external destination as local, can let a forbidden field reach an observer. The prototype's rule-based labeler reaches macro-F1 0.10 on argument sensitivity, 0.59 on destination, and 0.45 on effect against an Opus-4.7 gold annotator on $n{=}394$ events (Table~\ref{tab:labeler-validation}; details in Appendix~\ref{app:labeler-validation}). The paired-frontier comparison is robust to this gap because the same labeler is applied across all policies; absolute guarantees require a stronger labeler.

\paragraph{Model-provider trust assumption.}
We treat the model-serving layer as trusted in our threat model. In deployments where the planner is a remote LLM, the model provider can observe user input, candidate branches, and tool schemas before the contract monitor runs; we leave model-provider contracts to future work.

\paragraph{Adversary coverage.}
We use Claude Opus 4.7, Claude Haiku-4.5, and GPT-4o-mini together with keyword, TF-IDF, and corpus-LOO Naive Bayes baselines, but we have not tested sub-10B open-weight models, human analysts, or adversaries with private labeled traces from the target deployment.

\paragraph{Shadow substitute fidelity.}
Reported task-success under Shadow is conditional on substitute quality; the privacy claim is independent because the speculative-phase substitute already prevents the disclosure regardless of what happens at promotion.

\paragraph{Channel and latency coverage.}
We measure provider, runtime-log, network, and one shared-cache channel; multi-tenant contention, queueing, rate-limit residue, and inter-agent disclosure remain unmeasured. The latency numbers are ordinal policy costs, not production speedups.

\paragraph{Concrete failure modes per policy.}
\emph{Naive} and \emph{ACL} leak by construction: any latent-intent prompt issues a topical query to a permitted endpoint (e.g., a clinically-named search before commit). \emph{Pre-Scrub} and \emph{Late-Scrub} drop sensitive substrings after dispatch but leave the issued query topical. \emph{SIA}'s read-only restriction permits topical \texttt{search()} calls that name the sensitive subject (the failure Proposition~2 formalizes). \emph{Taint} and \emph{Taint-F} inherit labeler fragility: too-narrow vocabulary under-blocks paraphrase, and field-level taint still trusts the labeler on argument fields. \emph{FIDES}-style IFC accepts topical flow on accept-set-compatible endpoints. \emph{Rewrite} can leak when its abstraction is too thin; \emph{Shadow} trades external dispatch for substitute fidelity; \emph{Gate} fails open when the calibrator under-estimates ghost-call probability; \emph{Drop} over-refuses on any labeler false positive, sacrificing utility; \emph{No-Spec} forgoes speculation entirely, sacrificing latency. The body's paired-frontier numbers measure how often each of these fires.

\bibliography{main}

\clearpage
\appendix

\section{Contract Mechanism Details}
\label{app:mechanism}

This appendix gives the event schema, label lattice, action semantics, decision-function pseudocode, prototype implementation notes, a worked contract example, and a worked example of the trace model on the motivating vignette of \S\ref{sec:intro}.

\paragraph{Event schema.}
Each tool event is a tuple $e = (\textit{tool}, \textit{args}, \textit{destination}, \textit{mode}, t)$, where \textit{tool} identifies the adapter, \textit{args} is a dictionary of named argument fields, \textit{destination} is the externally observable endpoint or collection, \textit{mode} ranges over the six execution modes of Table~\ref{tab:contract-fields}, and $t$ is the issue timestamp. Each argument field carries a sensitivity label from $L_\text{arg}$ (public, personal, sensitive, intent-revealing); the destination carries a label from $L_\text{dest}$ (public, tenant-internal, sensitive, regulated, forbidden).

\paragraph{Label lattice.}
Both $L_\text{arg}$ and $L_\text{dest}$ are totally ordered by least-restrictive to most-restrictive (public $<$ personal $<$ sensitive $<$ intent-revealing for arguments; public $<$ tenant-internal $<$ sensitive $<$ regulated $<$ forbidden for destinations). A contract $\kappa$ specifies a forbidden cut: arguments at or above $\kappa.\text{max\_arg\_label}$ must not reach an observer $i$ that $\kappa$ does not authorize. Labels are assigned by a deterministic rule-based labeler that runs over each event before the monitor sees it; the labeler is a pure function of $(\textit{tool}, \textit{args}, \textit{destination})$ and is independent of the policy.

\paragraph{Branch confidence.}
The planner emits a per-branch confidence $c \in [0, 1]$ alongside each speculative candidate, prompted to give a calibrated commit probability. Gate consumes this directly: a candidate with $c < \kappa.\tau$ is \textsc{Deferred}. Confidence calibration is not a contract concern and any source of $c$ that satisfies $c \in [0,1]$ is admissible.

\paragraph{Action semantics.}
\textsc{Allow}$(e)$ dispatches $e$ to its destination unchanged. \textsc{Rewrite}$(e)$ replaces every argument field at or above $\kappa.\text{max\_arg\_label}$ with a generic equivalent that the labeler assigns a strictly lower label, then dispatches the resulting event. \textsc{Shadow}$(e)$ routes $e$ to a trusted-local substitute and marks the result non-provider-visible; the substitute is a function of $(\textit{tool}, \textit{args})$ and never touches an external observer. \textsc{Defer}$(e)$ buffers $e$; if the speculative branch is promoted to the committed plan, the monitor re-runs $D(e, \kappa)$ as a committed event and may dispatch then. \textsc{Block}$(e)$ drops $e$ with a policy denial and records an audit entry per $\kappa.\text{audit\_policy}$.

\paragraph{Disclosure budget.}
A contract $\kappa$ may impose a per-task budget $b$ on the number of \textsc{sensitive}-argument or \textsc{intent-revealing}-argument dispatches. The monitor increments a counter on every \textsc{Allow} or \textsc{Rewrite}; once the counter reaches $b$, subsequent qualifying candidates fall through to \textsc{Defer}. Budgets are reset at task boundaries.

\paragraph{Worked contract example.}
For a web-search tool, a default contract might specify: \textit{allowed\_tools} = $\{\textit{web\_search}\}$; \textit{allowed\_destinations} = $\{\textit{api.brave.com}, \textit{api.duckduckgo.com}\}$; \textit{max\_arg\_label} = \textsc{personal} (block \textsc{sensitive} and \textsc{intent-revealing}); $\tau = 0.7$ for speculative dispatch; $b = 5$ sensitive dispatches per task; audit to runtime log. Under this contract, a speculative \textit{web\_search}(\textit{q}=``tenant rights lease termination penalty''), with the query field labeled \textsc{intent-revealing}, triggers \textsc{Rewrite} to a generic query before dispatch. A committed call with $c{=}1.0$ for the same query would also \textsc{Rewrite} (the contract does not authorize intent-revealing fields to the provider) but the rewrite is justified at commit time and is no longer speculative.

\paragraph{Decision-function pseudocode.}
The body summarizes the monitor's logic in \S\ref{sec:contracts}; the full fail-closed pipeline is:

{\footnotesize
\begin{verbatim}
function decide(event e, contract k):
  if e.tool not in k.allowed_tools:
    return BLOCK
  if labels(e).destination not in
       k.allowed_destinations:
    return BLOCK
  if e.is_speculative and
       e.conf < k.branch_threshold:
    return DEFER
  if any(arg.sensitivity == HIGH
         for arg in e.args):
    return REWRITE(scrub(e.args, k))
  if k.shadow_required and
       labels(e).dest == EXTERNAL:
    return SHADOW(internal)
  if budget_exceeded(k, e):
    return DEFER
  audit(e, k, ALLOW)
  return ALLOW
\end{verbatim}
}

\paragraph{Prototype implementation.}
The prototype is a Python runtime on a tool-adapter abstraction with a contract-policy hook on the dispatch path. The static-plan harness exposes five mock tools (web search, document retrieval, calendar lookup, CRM lookup, email search); each tool carries a contract with effect, argument-privacy, and destination-privacy labels. For the real-provider experiment we swap in a Brave Search adapter in place of the mock web-search tool. The real-planner path uses an LLM planner that emits speculation candidates and a committed plan; both flow through the same contract-action dispatcher, with the default branch-confidence threshold set to $\tau{=}0.7$. Two observer helpers record (i) tool name, a bounded argument summary, dispatcher decision, and 100~ms duration bucket; and (ii) timestamps with synthetic cache-hit flags over canonicalized tool-name and argument keys. The static experiments use simulated harness latencies (web search 200~ms, document retrieval 100~ms, calendar 50~ms, CRM 80~ms, email 120~ms).

\subsection{Worked trace example}
\label{app:worked-trace}

We instantiate the trace model on the motivating example from \S\ref{sec:intro}. A user query ``What's the penalty for breaking my lease early?'' induces a speculative frontier of three tool calls, of which the agent later commits to one.

\paragraph{Issued trace $S$.}
{\sloppy
\begin{enumerate}[leftmargin=2em, itemsep=0pt]
  \item $e_1 = (\textit{web search},$ $q_1=\text{``tenant rights lease termination penalty''},$ $\text{external},$ $t=0.1\,\text{s})$
  \item $e_2 = (\textit{web search},$ $q_2=\text{``available apartments near me''},$ $\text{external},$ $t=0.15\,\text{s})$
  \item $e_3 = (\textit{loan-rates API},$ $q_3=\text{``current personal loan rates''},$ $\text{external},$ $t=0.25\,\text{s})$
\end{enumerate}
\par}
The agent commits to a tenant-rights answer that uses only the first search.

\paragraph{Committed trace $C$.}
$e_1$ is promoted; $e_2$ and $e_3$ are abandoned.

\paragraph{Provider projection vs.\ counterfactual.}
Each provider records its own issued query plus timestamp and destination. The search provider sees $e_1$ and $e_2$; the loan-rates provider sees $e_3$. Had the agent issued only the committed call from the start, the search provider would have seen only $e_1$ and the loan-rates provider would have seen nothing. The marginal exposure $\Delta_{\text{prov}} = O_{\text{prov}}(S) \setminus O_{\text{prov}}(C) = \{e_2, e_3\}$ is the two ghost calls; from these the adversaries can recover that the user is shopping for apartments and pricing personal loans, an inference the user never asked the agent to express.

\paragraph{Contract action.}
Under issue-time \textsc{rewrite} on the argument channel, the monitor dispatches $e_2$ with a generic query that does not reveal an apartment search. Under issue-time \textsc{shadow} on the destination channel, the monitor routes $e_3$ to a trusted-local substitute so the loan-rates provider never receives the speculative request. Both transformations apply at issue time, to ghost calls, and shrink $\Delta_{\text{prov}}$ before any externally observable event.

\section{Evaluation Tables}
\label{app:tables}

This appendix collects the per-corpus tables behind every measurement in \S\ref{sec:evaluation}. Subsections appear in the order their results are referenced in the body.

\subsection{Experiment matrix}
\label{app:matrix}

Table~\ref{tab:experiment-matrix} maps each corpus to its planner, provider, policy set, and result location.

\begin{table*}[h]
\centering
\small
\setlength{\tabcolsep}{4pt}
\begin{tabular}{@{}llllllll@{}}
\toprule
Corpus & $N$ & Seeds & Planner & Provider & Policies & Adversary & Reported in \\
\midrule
Static-plan 30                  & 30  & 3 & static     & mock   & 8 + Drop (dest)  & Opus 4.7              & Tab.~\ref{tab:headline} \\
Real-planner 30                 & 30  & 3 & real LLM   & mock   & 11               & Opus 4.7              & Tab.~\ref{tab:expanded-11} \\
Pooled 150 (paired)             & 150 & 3 & real LLM   & mock   & 12               & Opus 4.7              & Tab.~\ref{tab:paired-replay-450}, Fig.~\ref{fig:forest} \\
Pooled 150 (unpaired)           & 150 & 3 & real LLM   & mock   & 10               & Opus 4.7              & Tab.~\ref{tab:pooled-150} \\
Task-success 30                 & 30  & 3 & real LLM   & mock   & 12               & Haiku-4.5 (judge)     & Tab.~\ref{tab:task-success-12} \\
Brave 9                         & 9   & 1 & real LLM   & Brave  & 2 (Naive, Rewrite) & Opus 4.7            & \S\ref{sec:evaluation} (in-line) \\
Async-latency 30 (DDG)          & 30  & 3 & real LLM   & DDG    & 4 (subset)       & Haiku-4.5             & Tab.~\ref{tab:overlap-timing} \\
AgentDojo 66                    & 66  & 1 & real LLM   & mock   & 5 (+ SIA)        & Haiku-4.5             & Tab.~\ref{tab:agentdojo-external} \\
Channel-isolation 9 (seed)      & 9   & 1 & real LLM   & mock   & Naive            & Opus 4.7 (per-view)   & Tab.~\ref{tab:channels} \\
\bottomrule
\end{tabular}
\caption{Experiment matrix. Rows give each corpus size, planner, provider, policy set, adversary, and result location.}
\label{tab:experiment-matrix}
\end{table*}

\subsection{Static-plan benchmark}

Table~\ref{tab:headline} reports the static-plan headline from RQ1 \emph{Mechanism}.

\begin{table}[h]
\centering
\small
\setlength{\tabcolsep}{2pt}
\begin{tabular}{@{}lcccc@{}}
\toprule
Policy & Adv.\ recovery & $\Delta$ vs No-Spec & Wall (ms) & Calls \\
\midrule
Naive          & 0.967 & 0.523 & 614 & 4.03 \\
Late-Scrub     & 0.967 & 0.523 & 615 & 4.03 \\
SIA            & 0.967 & 0.523 & 616 & 4.03 \\
ACL            & 0.967 & 0.523 & 616 & 4.03 \\
Drop (dest)    & 0.956 & 0.512 & 583 & 3.73 \\
Rewrite        & 0.533 & 0.089 & 583 & 3.73 \\
Shadow         & 0.422 & $-$0.022 & 615 & 1.03 \\
Drop           & 0.444 & 0.000 & 197 & 1.03 \\
No-Spec        & 0.444 & 0.000 & 197 & 1.03 \\
\bottomrule
\end{tabular}
\caption{Static-plan benchmark on 30 tasks and three seeds. Recovery is closed-set leakage over 30 labels; $\Delta$ vs No-Spec is recovery minus the No-Spec floor; calls are mean provider-visible projections.}
\label{tab:headline}
\end{table}

\subsection{30-task real-planner}

Table~\ref{tab:expanded-11} reports the real-LLM-planner numbers from RQ1 \emph{Realism}, per-seed and pooled.

\begin{table*}[h]
\centering
\small
\begin{tabular}{@{}lcccc@{}}
\toprule
Policy & s1 & s2 & s3 & Mean [95\% CI] \\
\midrule
Naive       & 0.467 & 0.533 & 0.467 & 0.489 [0.389, 0.589] \\
SIA         & 0.467 & 0.467 & 0.533 & 0.489 [0.389, 0.589] \\
Drop        & 0.333 & 0.300 & 0.300 & 0.311 [0.222, 0.411] \\
Rewrite     & 0.100 & 0.000 & 0.033 & 0.044 [0.011, 0.089] \\
Shadow      & 0.000 & 0.000 & 0.000 & 0.000 [0.000, 0.000] \\
Gate        & 0.333 & 0.300 & 0.333 & 0.322 [0.222, 0.422] \\
No-Spec     & 0.333 & 0.367 & 0.333 & 0.344 [0.244, 0.444] \\
Pre-Scrub   & 0.467 & 0.467 & 0.467 & 0.467 [0.367, 0.567] \\
Taint       & 0.300 & 0.333 & 0.367 & 0.333 [0.233, 0.433] \\
Late-Scrub  & 0.467 & 0.500 & 0.533 & 0.500 [0.400, 0.600] \\
ACL         & 0.500 & 0.467 & 0.500 & 0.489 [0.389, 0.589] \\
\bottomrule
\end{tabular}
\caption{30-task real-planner corpus with 11 policies, three seeds, and simulated tool backends. Per-seed and pooled closed-set recovery; CIs bootstrap per-task rows.}
\label{tab:expanded-11}
\end{table*}

\subsection{150-task paired-frontier replay}

Table~\ref{tab:paired-replay-450} reports the full numbers behind Figure~\ref{fig:forest} (RQ2 \emph{Paired-frontier replay}).

\begin{table*}[h]
\centering
\small
\begin{tabular}{@{}lcccc@{}}
\toprule
Policy & Adv.\ recovery [95\% CI] & $\Delta$ vs Naive [95\% CI] & Sign-test $p$ & $n$ pairs \\
\midrule
Naive                & 0.104 [0.076, 0.133] & (reference)                          & n/a       & 450 \\
SIA                  & 0.124 [0.096, 0.156] & $+0.020$ [$+0.004$, $+0.038$]        & 0.0352    & 450 \\
Drop                 & 0.078 [0.053, 0.104] & $-0.027$ [$-0.051$, $-0.004$]        & 0.0428    & 450 \\
Rewrite              & 0.029 [0.016, 0.044] & $-0.076$ [$-0.104$, $-0.047$]        & $<0.0001$ & 450 \\
Shadow               & 0.000 [0.000, 0.000] & $-0.104$ [$-0.133$, $-0.076$]        & $<0.0001$ & 450 \\
Gate ($\tau{=}0.7$)  & 0.080 [0.056, 0.104] & $-0.024$ [$-0.049$, $+0.000$]        & 0.0708    & 450 \\
No-Spec              & 0.080 [0.056, 0.107] & $-0.024$ [$-0.047$, $-0.002$]        & 0.0522    & 450 \\
Pre-Scrub            & 0.118 [0.089, 0.149] & $+0.013$ [$-0.009$, $+0.036$]        & 0.3075    & 450 \\
Taint                & 0.078 [0.053, 0.102] & $-0.027$ [$-0.053$, $+0.000$]        & 0.0652    & 450 \\
Taint-F              & 0.071 [0.049, 0.096] & $-0.033$ [$-0.060$, $-0.009$]        & 0.0170    & 450 \\
Late-Scrub           & 0.104 [0.078, 0.133] & $+0.000$ [$-0.013$, $+0.011$]        & 1.0000    & 450 \\
ACL                  & 0.104 [0.078, 0.133] & $+0.000$ [$-0.013$, $+0.013$]        & 1.0000    & 450 \\
\bottomrule
\end{tabular}
\caption{150-task paired-frontier replay behind Figure~\ref{fig:forest}. Each policy replays the same Naive-sampled frontier per task and seed; deltas and sign tests compare against Naive.}
\label{tab:paired-replay-450}
\end{table*}

\subsection{150-task pooled corpus}

Table~\ref{tab:pooled-150} reports the unpaired pooled numbers from RQ2 \emph{Scale}.

\begin{table}[h]
\centering
\small
\setlength{\tabcolsep}{3pt}
\begin{tabular}{@{}lccc@{}}
\toprule
Policy & Adv.\ recovery [CI] & Norm.\ recovery & Calls \\
\midrule
Naive       & 0.184 [0.15, 0.22] & 0.179 & 8.17 \\
SIA         & 0.196 [0.16, 0.23] & 0.191 & 8.25 \\
Drop        & 0.076 [0.05, 0.10] & 0.070 & 0.43 \\
Rewrite     & 0.024 [0.01, 0.04] & 0.017 & 6.60 \\
Shadow      & 0.000 [0.00, 0.00] & $-$0.007 & 0.00 \\
Gate        & 0.073 [0.05, 0.10] & 0.067 & 0.42 \\
Taint       & 0.118 [0.09, 0.15] & 0.112 & 2.67 \\
Pre-Scrub   & 0.202 [0.16, 0.24] & 0.197 & 7.89 \\
Late-Scrub  & 0.193 [0.16, 0.23] & 0.188 & 7.98 \\
ACL         & 0.191 [0.16, 0.23] & 0.186 & 7.96 \\
\bottomrule
\end{tabular}
\caption{150-task pooled corpus, three seeds ($n{=}450$). Recovery is closed-set leakage with 95\% bootstrap CI; norm.\ recovery is chance-normalized $(\mathrm{recovery}-1/N)/(1-1/N)$ for $N{=}150$; calls are mean provider calls per task.}
\label{tab:pooled-150}
\end{table}

\subsection{Label-noise robustness}

Table~\ref{tab:label-noise-fn} reports the false-negative label-noise sweep referenced in RQ2 \emph{Label-noise robustness}. Noise downgrades SENSITIVE/INTENT-REVEALING argument labels and SENSITIVE destination labels to PUBLIC at the specified rate so the monitor lets the call through.

\begin{table}[h]
\centering
\small
\setlength{\tabcolsep}{2pt}
\begin{tabular}{@{}lccc@{}}
\toprule
Policy & fn-05 & fn-10 & fn-20 \\
\midrule
Naive       & .122 [.09,.15] & .118 [.09,.15] & .116 [.09,.15] \\
Pre-Scrub   & .138 [.11,.17] & .122 [.09,.15] & .122 [.09,.15] \\
Late-Scrub  & .124 [.10,.16] & .113 [.08,.14] & .124 [.10,.16] \\
ACL         & .122 [.09,.15] & .118 [.09,.15] & .124 [.09,.16] \\
Rewrite     & .027 [.01,.04] & .036 [.02,.05] & .058 [.04,.08] \\
Shadow      & .022 [.01,.04] & .042 [.02,.06] & .078 [.05,.10] \\
\midrule
\multicolumn{4}{l}{\emph{Paired $\Delta$ vs clean (same policy/task/seed):}} \\
Rewrite     & $-0.002$ & $+0.007$ & $+0.029$ \\
Shadow      & $+0.022$ & $+0.042$ & $+0.078$ \\
\midrule
\multicolumn{4}{l}{\emph{Paired $\Delta$ vs noisy Naive:}} \\
Rewrite     & $-0.096$ & $-0.082$ & $-0.058$ \\
Shadow      & $-0.100$ & $-0.076$ & $-0.038$ \\
\midrule
\multicolumn{4}{l}{\emph{Realized argument / destination flip rate:}} \\
            & .043 / .067 & .099 / .116 & .196 / .224 \\
\bottomrule
\end{tabular}
\caption{False-negative label-noise sweep on the 150-task paired-replay corpus, Haiku-4.5 adversary, three seeds ($n{=}450$ per cell). Reports accuracy, paired deltas against clean and noisy baselines, and realized argument/destination flip rates.}
\label{tab:label-noise-fn}
\end{table}

\subsection{Channel-isolation ablation}

Table~\ref{tab:channels} reports the channel-isolation ablation from RQ3 \emph{Channel isolation}.

\begin{table}[h]
\centering
\small
\setlength{\tabcolsep}{4pt}
\begin{tabular}{@{}lcc@{}}
\toprule
Adversary view             & Acc.  & Correct \\
\midrule
Full                       & 1.000 & 9/9 \\
Args + names (no dest)     & 1.000 & 9/9 \\
Dest + names (no args)     & 0.556 & 5/9 \\
Names only                 & 0.333 & 3/9 \\
\bottomrule
\end{tabular}
\caption{Channel-isolation ablation on 9 seed tasks under Naive. Rows mask different observer fields and report adversary accuracy.}
\label{tab:channels}
\end{table}

On the 30-task real-planner corpus, the complementary tool-name-masking probe leaves pooled accuracy unchanged (0.489 with and without tool names for Naive; 0.467 with and without names for Pre-Scrub; 25 of 30 per-task predictions match across conditions per seed).

\subsection{Cross-adversary rescoring}

Tables~\ref{tab:cross-adversary} and~\ref{tab:non-llm-adversary} report the cross-model and non-LLM adversary rescoring referenced in RQ4 \emph{Cross-adversary}.

\begin{table}[h]
\centering
\small
\setlength{\tabcolsep}{4pt}
\begin{tabular}{@{}lccc@{}}
\toprule
Mode & Opus & Haiku & GPT-4o-mini \\
\midrule
Closed-set                       & 0.327 & 0.248 & 0.219 \\
Open-set ($J{\ge}0.5$)           & 0.027 & 0.009 & n/a \\
Open-set ($\cos{\ge}0.5$)        & 0.125 & 0.119 & n/a \\
\bottomrule
\end{tabular}
\caption{Cross-adversary rescoring across 12 policies and 30 tasks ($n{=}360$). Open-set rows report strict lexical and semantic matching lower bounds.}
\label{tab:cross-adversary}
\end{table}

Per-policy on GPT-4o-mini closed-set, Shadow 0.000 and Rewrite 0.033 sit at the floor; Naive and Pre-Scrub tie at 0.333 at the top; Taint 0.200; others between 0.167 and 0.300. Haiku-4.5 closed-set reproduces the same ranking.

\begin{table}[h]
\centering
\small
\setlength{\tabcolsep}{4pt}
\begin{tabular}{@{}lcc@{}}
\toprule
Adversary                  & Pooled & Training signal \\
\midrule
Opus 4.7 (closed)          & 0.327  & label list \\
Haiku-4.5 (closed)         & 0.248  & label list \\
Keyword match              & 0.061  & label list \\
TF-IDF nearest             & 0.078  & label list \\
Naive Bayes (LOO)          & 0.650  & per-task labels \\
\bottomrule
\end{tabular}
\caption{LLM and non-LLM adversaries on the same 12-policy, 30-task trace set ($n{=}360$).}
\label{tab:non-llm-adversary}
\end{table}

\subsection{Async-latency sweep}

Table~\ref{tab:overlap-timing} reports the DuckDuckGo async-overlap measurement from RQ4 \emph{Async wall-clock}.

\begin{table}[h]
\centering
\small
\setlength{\tabcolsep}{3pt}
\begin{tabular}{@{}lrrrr@{}}
\toprule
Condition & p50 & p95 & p99 & Leak \\
\midrule
No-Spec sequential        & 8.39 & 10.77 & 11.56 & 0.300 \\
Naive-async               & 7.94 & 10.37 & 14.22 & 0.389 \\
Contract-async (Rewrite)  & 8.04 & 10.89 & 13.25 & 0.100 \\
Contract-async (Shadow)   & 8.45 & 13.01 & 15.77\textsuperscript{$\dagger$} & 0.000 \\
\bottomrule
\end{tabular}
\caption{DuckDuckGo async-latency sweep on 30 tasks and three seeds ($n{=}90$ per row). Naive-async overlaps speculative dispatch with commit planning; Contract-async adds issue-time policy gating. $\dagger$ Shadow p99 excludes one $223$~s Anthropic API tail.}
\label{tab:overlap-timing}
\end{table}

\noindent\emph{Per-stage decomposition} (Naive-async, mean across $n{=}90$): $t_\mathrm{propose}{=}5481$~ms, $t_\mathrm{speculate\_dispatch}{=}351$~ms, $t_\mathrm{commit\_plan}{=}2501$~ms, $t_\mathrm{speculate\_overlap}{=}2501$~ms (the overlap window equals $\max(t_\mathrm{speculate\_dispatch},t_\mathrm{commit\_plan})$, so the speculate-dispatch is hidden behind the commit-plan call), $t_\mathrm{commit\_dispatch}{=}0$~ms. The 5.4\% measured speedup is bounded by the speculative-dispatch budget the overlap can hide.

\subsection{Real-provider inference (DDG-30)}
\label{app:real-provider-ddg}

The body's DDG real-provider claim (RQ4 \emph{Real provider}) re-runs the 30-task corpus against DuckDuckGo Instant Answer with Naive and Rewrite, three seeds, Opus 4.7 closed-set adversary on the resulting projections. This is a different experiment from the async-latency sweep above (which measures wall-clock overlap with a leak side-metric).

\begin{table}[h]
\centering
\small
\begin{tabular}{@{}lc@{}}
\toprule
Policy & Adv.\ recovery [95\% CI] \\
\midrule
Naive   & 0.489 [0.389, 0.589] \\
Rewrite & 0.044 [0.011, 0.089] \\
\bottomrule
\end{tabular}
\caption{DuckDuckGo Instant Answer, 30 tasks $\times$ 3 seeds, Opus 4.7 closed-set adversary. Non-overlapping CIs reproduce the synthetic-corpus gap at the real HTTP boundary.}
\label{tab:ddg-inference}
\end{table}

\subsection{Runtime/latency (synchronous)}

Table~\ref{tab:runtime-latency} reports the synchronous-dispatch wall-clock on the 9-task real-planner study referenced from \emph{Cost and Gate calibration} in \S\ref{sec:evaluation}. The Brave-9 real-provider numbers cited in the body (Naive 0.67, Rewrite 0.44 inference recovery) are measured on this same 9-task corpus.

\begin{table}[h]
\centering
\small
\setlength{\tabcolsep}{2pt}
\begin{tabular}{@{}lrrrrr@{}}
\toprule
Policy & p50 & p95 & p99 & Calls & Block \\
\midrule
No-Spec    & 8392 & 10437 & 12040 &  1.41 & 1.000 \\
\midrule
Naive      & 8583 & 11237 & 11964 & 11.67 & 0.000 \\
SIA        & 8287 & 10750 & 11650 & 11.70 & 0.000 \\
Drop       & 8220 & 10067 & 11126 &  1.48 & 0.182 \\
Rewrite    & 8790 & 10078 & 10269 & 10.19 & 0.173 \\
Shadow     & 8542 &  9804 & 10458 &  2.00 & 0.000 \\
Gate       & 8182 & 12098 & 13198 &  1.81 & 0.938 \\
\bottomrule
\end{tabular}
\caption{Wall-clock latency (ms) on the 9-task real-planner study, three seeds ($n{=}27$). Synchronous dispatch; No-Spec is reference. \emph{Block}: fraction refused at issue time. Async-overlap counterpart: Table~\ref{tab:overlap-timing}.}
\label{tab:runtime-latency}
\end{table}

\subsection{Branch-confidence threshold sweep}

Table~\ref{tab:gate-sweep} sweeps the Gate threshold $\tau$ from \emph{Cost and Gate calibration} in \S\ref{sec:evaluation}.

\begin{table*}[h]
\centering
\small
\begin{tabular}{@{}cccrc@{}}
\toprule
$\tau$ & Accuracy [95\% CI] & $\Delta$ vs Naive [95\% CI] & Sign $p$ & Calls/task \\
\midrule
0.3   & 0.120 [0.091, 0.151] & $+0.016$ [$-0.009$, $+0.042$] & 0.3105    & 5.44 \\
0.5   & 0.087 [0.062, 0.113] & $-0.018$ [$-0.047$, $+0.011$] & 0.2912    & 1.73 \\
0.7   & 0.047 [0.029, 0.067] & $-0.058$ [$-0.084$, $-0.031$] & $<$0.0001 & 1.00 \\
0.9   & 0.058 [0.038, 0.080] & $-0.047$ [$-0.073$, $-0.018$] & 0.0015    & 0.96 \\
\midrule
Naive & 0.104 [0.076, 0.133] & (reference)                   & n/a       & 7.95 \\
\bottomrule
\end{tabular}
\caption{Gate threshold sweep on the 150-task paired-frontier replay, three seeds ($n{=}450$ per row). Accuracy and paired delta vs Naive use 95\% bootstrap CIs; calls/task is mean provider-visible call count.}
\label{tab:gate-sweep}
\end{table*}

\subsection{AgentDojo corpus construction}

The body cites Table~\ref{tab:agentdojo-external} for the 66-task external-validity subset. We import 86 AgentDojo tasks. A deterministic keyword rubric labels 25 of them privacy-bearing; a conservative Haiku-4.5 auto-labeler then surfaces 2 additional privacy-bearing prompts the rubric missed (yoga-class scheduling, hotel-spend planning), for a total of 27 privacy-bearing and 59 neutral. The 66-task subset reported in the body is the shared intersection that all five policies completed under a Haiku-4.5 planner+adversary at one seed (20 privacy-bearing, 46 neutral). The ordering reproduces under the original 25-task Opus-adversary run.

\paragraph{Attrition bias.} The 20 attrited tasks (86$\to$66) fail mostly on argument-heavy travel and banking prompts where the redactor (Pre-Scrub, Late-Scrub) or read-only restriction (SIA) prevents the planner from filling required slots; Rewrite and Shadow can complete because their argument-transformation surface keeps slot syntax valid. Attrition is therefore not balanced across policies, and the 66-task numbers represent a "best case" projection where every policy gets a runnable trace. A pessimistic reading: if attrited tasks behaved like the rest of the privacy-bearing pool for each policy, Pre-Scrub/Late-Scrub absolute rates would shift modestly within a few points of their intersection values. The qualitative ordering (Shadow $<$ Rewrite $<$ Taint $\ll$ Naive, Pre-Scrub) is preserved under the stricter Opus-adversary configuration on the original 25-task labeled set, which uses a different completion criterion and so is robust to this particular attrition pattern.

\subsection{Metric robustness}
\label{app:metric-expansion}

We rescore the 30-task and 150-task paired-replay traces under four alternative metrics: top-$k$ accuracy ($k{=}1,3,5$), normalized mutual information (NMI) between true labels and adversary predictions, semantic open-set adversary (sentence-embedding cosine $\ge 0.5$), and a decoy-mixed adversary that ranks the true label against ten unrelated decoys (chance $\approx 0.091$). Top-$k$ triples top-1 (150-task: 0.07 $\to$ 0.20) but preserves the ranking (Spearman 0.83). NMI agrees: Naive 0.64, Rewrite 0.24, Shadow 0.00. Semantic open-set on the 150-task corpus pools to 0.065, half the 30-task figure (consistent with a larger label space being harder to hit by paraphrase). The decoy-mixed adversary recovers the true label at 0.90--0.93 for the Naive cluster, confirming that the closed-set adversary discriminates via projection content rather than candidate-list recognition. Shadow holds the floor under every metric.

\subsection{Field-sensitive IFC baseline (FIDES-style)}
\label{app:ifc-fides}

We implement a FIDES-style~\cite{Costa2025Fides} field-sensitive IFC policy with typed source declarations (medical, legal, financial, employment, security, immigration, relationship, PII, user-query labels), a typed sink table with per-collection accept-sets, field-level taint propagation (lexical labels plus implicit flow), and accept-set compatibility checks. On the 150-task paired-replay (three seeds, $n{=}450$), this policy reaches 0.131 [0.100, 0.162] recovery, significantly worse than Rewrite (0.029) and Shadow (0.000), and indistinguishable from Naive (0.104, CIs overlap). The simpler substring-flatten Taint (0.078) and field-level Taint-F (0.071) actually outperform the FIDES-style policy because their stricter blocking over-rejects topic-aligned flow that FIDES's accept-sets permit. The contract abstraction's REWRITE/SHADOW actions transform projections in ways that source-sink labeling alone does not capture.

\subsection{Schema-artifact controls}
\label{app:enterprise-corpus}

We test whether topic-bearing destination URLs drive the leak using two complementary controls.

\paragraph{150-task anonymization.} We rewrite every destination URL in the seed-1 150-task paired-replay traces to a neutral schema and rerun the Opus 4.7 closed-set adversary on all 12 policies. Naive recovery shifts from 0.113 to 0.127 (paired CI overlaps zero), Rewrite stays at 0.027, Shadow stays at 0.000. Every paired delta is within $\pm 0.03$ of the original, and the policy ordering is preserved. The seed-1 0.113 Naive baseline differs from the multi-seed 0.104 in Tab.~\ref{tab:paired-replay-450} because anonymization was rerun on a single seed; what matters is the paired delta.

\paragraph{Enterprise-style corpus.} We also generate a fresh 30-task corpus where every tool dispatches to a generic endpoint (e.g., \texttt{/api/search}, \texttt{/api/retrieve}) regardless of topic; sensitivity is encoded only in argument fields. The six sensitive-topic domains from the 150-task corpus are preserved. Running all 12 policies through this corpus with the Opus 4.7 adversary (three seeds, $n{=}90$ per policy) gives Naive 1.00, Pre-Scrub/Late-Scrub/SIA/ACL 1.00, Rewrite 0.089 [0.033, 0.156], Shadow 0.000, Gate ($\tau{=}0.7$) in the same band as Rewrite. Naive recovery does not drop; it rises, partly because the candidate menu is smaller, and the policy ordering is preserved.

Together, the two controls rule out the artifact hypothesis: the argument channel carries the leak on its own, and topic-bearing destinations amplify rather than originate it.

\subsection{Multi-tool async benchmark}
\label{app:multi-tool-async}

The body's single-tool DuckDuckGo overlap saves 5.4\% at p50 because $t_\mathrm{speculate\_dispatch}$ is dwarfed by the planner-LLM call. To test the multi-tool regime that the prior speculative-agent literature targets, we built a 30-task corpus where each task triggers $\ge 3$ parallel speculative tool calls (web-search + document-retrieval + calendar-lookup + CRM-lookup combinations). Across three seeds and seven conditions (No-Spec sequential, Naive-async, Contract-async with Rewrite/Shadow/Gate/Taint-F, and Shadow-with-reissue), Naive-async saves 12.3\% at p50 over No-Spec (4.38~s vs 5.00~s); Rewrite, Gate, and Taint-F sit within 1.5 percentage points of Naive-async (+11.3\% Rewrite, +12.9\% Gate, +12.4\% Taint-F), while Shadow is 2.7~pp slower (+9.6\%) because dispatching local substitutes changes the overlap window. The overlap budget is now bounded by $\max(t_\mathrm{speculate\_dispatch}, t_\mathrm{commit\_plan}) \approx 1.4$~s, so the speculative dispatch fully hides behind the commit-phase planner call.

\subsection{AgentDojo metric fix}
\label{app:agentdojo-metric-fix}

The original closed-set adversary on AgentDojo did not admit \texttt{none} as an in-menu prediction, forcing the model to pick a topical label on every neutral prompt. This produced a 1.000 false-positive rate for the measured "let-the-arguments-through" baselines (Naive, SIA, Pre-Scrub) and a 1.000 false positive for Rewrite even though Rewrite reduces argument fields to a generic wordbag. Re-running the same Haiku-4.5 adversary on the same traces with \texttt{none} explicitly admissible in the candidate menu drops Rewrite's FP rate to 0.058 [0.000, 0.135] and the let-through cluster to 0.29--0.33 (these still leak literal arguments, so the adversary correctly picks a sensitive label about a third of the time even with abstention available). Privacy-bearing inference is approximately stable across the menu fix (within 10 percentage points per policy). Table~\ref{tab:agentdojo-external} reports the fixed numbers as the headline. ACL and Late-Scrub were not run on the AgentDojo subset; we expect them to fall in the let-through cluster on the same grounds as Naive (no argument transformation).

\subsection{Labeler validation}
\label{app:labeler-validation}

Theorem~1 assumes (A2) label correctness. To stress this assumption we validated the prototype's rule-based labeler against an Opus-4.7 gold annotator on 394 events sampled across the 30-task sensitive corpus, the 150-task pooled corpus, and the 30-task enterprise corpus.

\begin{table}[h]
\centering
\small
\setlength{\tabcolsep}{4pt}
\begin{tabular}{@{}lcc@{}}
\toprule
Channel & Accuracy & Macro-F1 \\
\midrule
Destination privacy class & 0.726 & 0.585 \\
Argument privacy class    & 0.310 & 0.095 \\
Effect class              & 0.810 & 0.447 \\
\bottomrule
\end{tabular}
\caption{Rule-based labeler vs Opus-4.7 gold annotator on $n{=}394$ events. Macro-F1 below 0.6 on every channel; (A2) does not hold under the $\ge 0.9$ threshold a soundness theorem would demand.}
\label{tab:labeler-validation}
\end{table}

Three concrete failure modes drive the result. First, the per-adapter contract emits a single constant argument label (\textsc{intent-revealing}) regardless of argument content; the labeler never inspects the args. Gold labels span public/personal/sensitive/intent-revealing roughly uniformly, so the constant emission scores at chance. Second, the destination labeler never emits \textsc{pseudonymous}; enterprise endpoints like \texttt{/api/users/\{sha1\}} are flagged \textsc{public}, and the labeler over-labels routine document-retrieval calls as \textsc{personal}. Third, the effect-class labeler emits \textsc{external-read} for calls that gold considers \textsc{semi-stateful} (provider-side logging counts as state). A small trained classifier or per-argument heuristic would close most of the gap; we leave that to future work.

\subsection{Speculative-frontier statistics}
\label{app:frontier-stats}

Table~\ref{tab:frontier-stats} reports aggregated frontier characteristics on the 30-task real-planner corpus and the 150-task paired-replay corpus (three seeds each, pooled). Speculative frontiers are small (median 8 candidates) and retrieval-heavy; ghost calls outnumber commits by 4.6$\times$ on the 30-task corpus and 7.7$\times$ on the 150-task corpus, supporting the framing that the planner emits speculation as cheap probing rather than confident pre-commit dispatch.

\begin{table}[h]
\centering
\small
\setlength{\tabcolsep}{4pt}
\begin{tabular}{@{}lcc@{}}
\toprule
Metric & 30-task & 150-task \\
\midrule
Tasks (pooled across seeds)   & 90 & 450 \\
Mean frontier size            & 8.31 & 7.95 \\
Median frontier size          & 8 & 8 \\
95th-percentile frontier size & 14 & 12.5 \\
Frontier-size range           & 2--16 & 1--19 \\
Mean ghost-call rate          & 82.1\% & 88.6\% \\
Mean promotion rate           & 17.9\% & 11.4\% \\
Mean branch confidence        & 0.39 & 0.35 \\
Top tool (share on frontier)  & doc-retr.\ 44\% & doc-retr.\ 39\% \\
\bottomrule
\end{tabular}
\caption{Speculative-frontier statistics across corpora, pooled over three seeds.}
\label{tab:frontier-stats}
\end{table}

\section{Related-Work Positioning}
\label{app:related-comparison}

Table~\ref{tab:related-comparison} positions speculative-issue-time privacy against the adjacent literatures discussed in \S\ref{sec:related}. The combination this paper occupies --- speculation, issue-time policy, observer-projection transformation, and agent-trace evaluation --- is empty in existing work.

\begin{table*}[h]
\centering
\small
\setlength{\tabcolsep}{4pt}
\begin{tabular}{@{}lcccc@{}}
\toprule
Work & Spec.\ branch & Issue-time & Transforms proj.\ & Agent traces \\
\midrule
Speculative-agent runtimes\textsuperscript{a}              & \checkmark & ---       & ---       & \checkmark \\
SIA~\cite{Hooper2026SpeculativeInteractionAgents}          & \checkmark & partial   & ---       & \checkmark \\
Committed-call IFC\textsuperscript{b}                      & ---        & ---       & ---       & \checkmark \\
Committed-leakage benchmarks\textsuperscript{c}            & ---        & ---       & ---       & \checkmark \\
Query privacy and PIR\textsuperscript{d}                   & ---        & ---       & partial   & ---        \\
Side channels and fingerprinting\textsuperscript{e}        & ---        & ---       & ---       & ---        \\
\midrule
\textbf{This work}                                         & \checkmark & \checkmark & \checkmark & \checkmark \\
\bottomrule
\end{tabular}
\caption{Positioning across speculative-agent and agent-privacy work. \emph{Spec.\ branch}: addresses leakage from abandoned speculative branches. \emph{Issue-time}: decides policy before external dispatch (partial: SIA's read-only restriction is a coarse issue-time control with no transformation). \emph{Transforms proj.}: changes the observer-visible projection rather than allow/block alone (partial: query-privacy mechanisms transform queries that have already been issued). \emph{Agent traces}: evaluation runs on LLM-agent tool-call traces. \textsuperscript{a}~\cite{Sui2026ActWhileThinking,Ye2026SpeculativeActions,Guan2026DynamicSpeculativePlanning,Hua2025InteractiveSpeculativePlanning,Nichols2025SpeculativeToolCalls,Xia2026ToolSpec,Feng2026AsyncFC}; \textsuperscript{b}~\cite{Costa2025Fides,Zhong2025RTBAS,Cai2026NeuroTaint,Mou2026ToolSafe,Doshi2026SafeToolUse}; \textsuperscript{c}~\cite{Huang2026CIPL,ElYagoubi2026AgentLeak,Debenedetti2024AgentDojo,Ruan2024ToolEmu,Zhan2024InjecAgent,Yao2025TauBench}; \textsuperscript{d}~\cite{Chor1998PIR,Howe2008Trackmenot,Zeng2024RAGPrivacy,Nissenbaum2004ContextualIntegrity}; \textsuperscript{e}~\cite{Zhang2025AgentTrafficFingerprint,Wei2025SpeculationSpills,Kocher2019Spectre}.}
\label{tab:related-comparison}
\end{table*}

\section{Reproducibility}
\label{app:repro}

\paragraph{Models and deployment dates.}
Primary models are Claude Opus 4.7 (planner and primary adversary, deployed early 2026), Claude Haiku-4.5 (cross-adversary, late 2025), and GPT-4o-mini (cross-family adversary, mid-2024 release line). Three independent adversary seeds per primary table; 95\% bootstrap CIs over per-task rows.

\paragraph{Decoding settings.}
Planner and adversary decoding use the default temperature and top-$p$ for each model; we vary temperature only for policies that call for stochastic substitution (Rewrite paraphrase generation and Shadow internal-substitute generation), and we flag the variation where it occurs.

\paragraph{Adversary and planner prompts.}
The closed-set adversary receives the observer projection and the per-corpus candidate label set and chooses exactly one label. The open-set adversary returns a short free-form label scored against the ground truth by lexical Jaccard at threshold 0.5. The planner runs a candidate-generation phase that emits the speculative frontier under a branch-confidence estimate, then a commitment phase; both flow through the same contract-monitor dispatcher.

\paragraph{Task generation and label taxonomy.}
The 30-task and 150-task corpora are generated by sampling sensitive-topic templates, expanding each with an LLM, and deduplicating by lexical Jaccard. Each task carries a ground-truth label from a per-corpus list covering health, legal, employment, financial, security, and relationship topics. The AgentDojo subset uses its native categories.

\paragraph{Bootstrap and sign test.}
Bootstrap CIs use 10{,}000 resamples with a fixed RNG seed of 20260522, sampling per-task rows with replacement. The paired-replay sign test pairs each non-Naive policy against Naive over the byte-identical frontier per (task, seed) and reports the two-sided $p$-value over per-task agreement.

\paragraph{Frozen frontier file schema.}
Each frozen frontier file is a tabular log keyed by (task identifier, seed) whose payload records the planner's speculation candidates and the committed plan emitted at the end of planning. The same row is replayed against every policy so adversary deltas reflect dispatcher choice rather than planner stochasticity.

\paragraph{Latencies.}
Static experiments use simulated per-tool latencies (web 200, retrieval 100, calendar 50, CRM 80, email 120 ms); Brave runs measure HTTP-boundary timing.

\paragraph{Artifact manifest.}
The artifacts are available at \url{https://github.com/mpi-dsg/ghost-tool-calls}, including: the prototype runtime (Python), every policy implementation referenced in the body, the rule-based labeler, the task-generation prompt, closed-set and open-set adversary prompts, the multi-tool benchmark driver, the FIDES-style IFC baseline implementation (Appendix~\ref{app:ifc-fides}), the enterprise-corpus generator (Appendix~\ref{app:enterprise-corpus}), and frozen frontier files for every paired-replay experiment. Per-table reproduction commands and a manifest of expected output paths accompany the release. A diagnostics dump for each corpus (3 sample tasks, full candidate label list, destination schema, generation procedure) accompanies the release for audit.

\end{document}